\documentclass[a4paper]{jpconf}
\usepackage{graphicx}
\usepackage{amsmath,amsfonts,amssymb,bm}

\newcommand{\beq}{\begin{equation}} \newcommand{\eeq}{\end{equation}}
\newcommand{\bes}{\begin{split}} \newcommand{\ees}{\end{split}} 
\newcommand{\bea}{\begin{eqnarray}} \newcommand{\eea}{\end{eqnarray}}

\def\s{\sigma}

\def\us{{\underline{\sigma}}}
\def\ubs{{\underline{\boldsymbol{\sigma}}}}

\def\hsix{{\widehat{\sigma}_i^x}}

\def\hsiz{{\widehat{\sigma}_i^z}}
\def\hH{\widehat{H}}

\def\la{\langle}
\def\ra{\rangle}

\def\T{\mathcal{T}}

\def\bs{{\boldsymbol{\sigma}}}

\def\dd{{\rm d}}

\def\tm{\widetilde{m}}

\begin{document}

\title{Thermal, quantum and simulated quantum annealing: analytical comparisons for simple models}

\author{V Bapst and G Semerjian}

\address{LPTENS, Unit\'e Mixte de Recherche (UMR 8549) du CNRS et
  de l'ENS, associ\'ee \`a l'UPMC Univ Paris 06, 24 Rue Lhomond, 75231
  Paris Cedex 05, France.}

\ead{victor.bapst@polytechnique.edu,guilhem@lpt.ens.fr}

\begin{abstract}
We study various annealing dynamics, both classical and quantum, for simple mean-field models and explain how to describe their behavior in the thermodynamic limit in terms of differential equations. In particular we emphasize the differences between quantum annealing (i.e.\ evolution with Schr\"odinger equation) and simulated quantum annealing (i.e.\ annealing of a Quantum Monte Carlo simulation).
\end{abstract}

\section{Introduction}

Combinatorial optimization is a branch of computer science concerned with the task of finding the minimum of a cost function $C(\us)$ of discrete variables $\us=(\s_1,\dots,\s_N)$. The computational complexity theory~\cite{Papadimitriou94} classifies the difficulty of such a task, depending on the problem type (the ``shape'' of the function), in terms of the time and memory requirements of algorithms that find optimal configurations of the variables (or approximations with a controlled accuracy). If the cost function is viewed as an energy, optimal configurations correspond to groundstates and this analogy leads to physics-inspired annealing procedures aiming at discovering the groundstate of a given system. Thermal (simulated) annealing~\cite{sa} in particular exploits thermal fluctuations to achieve this goal: in a Monte Carlo numerical experiment a fictitious temperature can be slowly decreased down to very low temperatures. If this decrease is slow enough the system remains close to equilibrium at all times, and at the end of this annealing, where the temperature is very low, the typical configurations encountered will be the groundstates of the system. Instead of thermal fluctuations, or in addition to them, one can envision the use of quantum fluctuations to accelerate the exploration of the configuration space towards the discovery of the groundstate. This idea appeared with several variants and names~\cite{qa_first,qa_second,qa,Aeppli99,qaa}, in particular quantum annealing and quantum adiabatic algorithm (see~\cite{qa_review_santoro,qa_review_das,review_Nishimori,long} for reviews). In one of these variants one considers the evolution of an isolated system at zero temperature, thus described by the Schr\"odinger equation, with a quantum Hamiltonian interpolating during the course of the evolution between a simple Hamiltonian with an easy to prepare groundstate (typically a transverse field) and the Hamiltonian encoding the cost function one wants to minimize. If this interpolation is slow enough (i.e.\ if the adiabaticity condition is fulfilled), a system initially prepared in the groundstate of the simple Hamiltonian will end up in the sought-for groundstate of the cost function. This procedure, that we shall call quantum annealing in the following, could be implemented in an ideal quantum computer (with a perfect coherence), but its simulation on a classical computer is of course very costly (the dimension of the quantum Hilbert space growing exponentially with the number of variables). Another variant, that we shall call simulated quantum annealing in the following (also known as Path-Integral Monte Carlo Quantum Annealing in~\cite{qa_review_santoro,review_Nishimori} or Quantum Monte Carlo Annealing in~\cite{qa_review_das}), amounts to perform a Quantum Monte Carlo numerical experiment, based on an imaginary-time path integral representation of the quantum partition function, at positive temperature, in which some parameters, in particular the transverse field, are slowly modified during the simulation. Recent claims that such a procedure can be implemented experimentally with a rather large number of variables can be found in~\cite{dwave_troyer}.

A satisfactory understanding of these various annealing procedures would require being able to answer at least the following two questions: i) how long should be the annealing to reach the groundstate? ii) if the annealing is too fast, what is the residual energy (i.e.\ the difference between the energy of the configuration reached at the end of the annealing and the one of the groundstate) as a function of the annealing time? These questions, and in particular the behavior of these quantities in the thermodynamic limit, have notably been studied in the context of mean-field spin-glasses~\cite{book_Marc_Andrea}, that are relevant as typical instances of random combinatorial optimization problems~\cite{CheesemanKanefsky91}. However, even in this mean-field setting, only partial answers to questions i) and ii) are known. Consider for instance the thermal annealing case; mean-field spin-glasses do exhibit a so-called dynamic transition temperature $T_{\rm d}$, below which metastable pure states proliferate and drive the system out-of-equilibrium if the annealing is not slow enough to cross the free energy barriers between the pure states~\cite{CuKu93}. In many of these models these barriers are extensive, hence the time required for an annealing to bring the system in its groundstate is exponentially large in the size of the system. Even if the computation of $T_{\rm d}$ is now well-understood, the rate of extensive growth of the barriers, hence the answer to question i), is generically unkown. Question ii) can be more easily adressed in a particular limit, i.e.\ for annealing durations which go to infinity after the thermodynamic limit has been taken, see in particular~\cite{ZdKr10,FrPa13,KrZd_japan} for recent works on this question with the so-called state following method.

The current knowledge for the analysis of the quantum annealing procedure is similar and slightly less advanced: from the equilibrium analysis of quantum mean-field spin glasses, via the replica method for fully-connected models or the cavity one for diluted models (see~\cite{long} for a recent review of these works), the location of the quantum phase transition where the gap between the groundstate and the first excited one closes in the thermodynamic limit can be computed (similarly to $T_{\rm d}$ for the classical case), yet the quantitative behavior of the gap is known analytically only in some simple cases~\cite{jorg08,farhi10,BaSe12}. The equivalent of the state following method in the quantum case has not yet been developed. For what concerns the simulated quantum annealing, the dynamic transition of the path-integral model representing the quantum problem at finite temperature has been computed; as demonstrated in~\cite{long} this is the point at which a QMC annealing falls out-of-equilibrium, but again no prediction is available for its long (but subexponential) time behaviour. Let us also mention the existence of worst-case bounds on the annealing schedules of both quantum and simulated quantum annealing that ensure the convergence to the groundstate in an infinite time (see~\cite{review_Nishimori} for a review of these results) and of several numerical experiments comparing the efficiency of the various annealing procedures (for instance~\cite{santoro02,StSaTo06}, and~\cite{qa_review_santoro,qa_review_das} and references therein).

The strategy followed in this paper consists in studying simple models on which these questions can be investigated analytically, namely non-disordered fully-connected models. The thermal annealing of these models is rather simple to analyze, see~\cite{GrWeLa66} for early results. We previously considered in great details the quantum annealing procedure (i.e.\ the real-time Schr\"odinger evolution)~\cite{BaSe12}, here we show that the simulated quantum annealing case is also tractable (which was previously discovered in~\cite{Inoue}) and we emphasize the difference between the last two dynamics.

The rest of this paper is organized as follows. In Sec.~\ref{sec_definitions} we give explicit definitions of the various annealing dynamics and of the models to be studied. Then in Sec.~\ref{sec_evolution_eq} we present the evolution equations that govern the dynamics in the thermodynamic limit, and study their long-time behavior; this unveils a crucial difference in the behavior of the quantum annealing and the simulated quantum annealing (even at very low temperature). We finally come to our conclusions in Sec.~\ref{sec_conclusions}, sketching some extensions of these results to more complicated models and other time regimes. For reason of space limitations we refrain from giving details of the computations that shall be presented elsewhere~\cite{BaSe13long}.

\section{Definitions}
\label{sec_definitions}
\subsection{Various annealing dynamics}

We consider models whose classical configurations are given by the values of $N$ Ising spins, denoted $\us=(\s_1,\dots,\s_N)$. In the quantum case we consider the $2^N$-dimensional Hilbert space spanned by the vectors denoted $\{|\us\ra \}$, that form the so-called computational basis. We write $\hsiz$ and $\hsix$ for the Pauli matrices acting on the $i$-th spin according to $\hsiz |\us\ra = \s_i |\us\ra $ and $\hsix |\us\ra = | \us^{(i)}\ra $, where $\us^{(i)}$ is the configuration obtained from $\us$ by flipping the $i$-th spin.

In the various annealing procedures we denote $\T$ the total physical time of the evolution, which is parametrized by a reduced time parameter $s\in[0,1]$. All parameters are allowed to evolve with $s$, i.e.\ $\beta(s)$ is the time varying inverse temperature, and in the most general case we consider a time-dependent quantum Hamiltonian written as
\beq
\hH(s) = \sum_\us E(\us;s) |\us\ra \la \us | - \Gamma(s) \sum_{i=1}^N \hsix \ .
\eeq
The ``classical'' (i.e.\ diagonal in the computational basis) energy $E(\us;s)$ can a priori vary in time, as well as the intensity $\Gamma(s)$ of the transverse field.

Let us now define the dynamics to be studied.

\subsubsection{Thermal annealing}

This first case is purely classical ($\Gamma(s)=0$), with a dynamics made of a Markov stochastic process in the configuration space, described by a continuous time master equation for the probability $P(\us;s)$ to observe the system in the configuration $\us$ at (reduced) time $s$:
\beq
\frac{1}{\T} \frac{\dd  }{\dd s} P(\us;s) = \sum_{\us'} W(\us,\us';s) P(\us';s) \ .
\label{eq_master_thermal}
\eeq
The jump rates from configuration $\us'$ to $\us$ (with diagonal terms chosen to ensure the conservation of probabilities) are assumed to verify the detailed balance condition with respect to the instantaneous Gibbs-Boltzmann distribution, i.e.
\beq
W(\us,\us';s) e^{-\beta(s) E(\us';s)} = W(\us',\us;s) e^{-\beta(s) E(\us;s)} \ .
\eeq
In the usual form of simulated annealing $E(\us;s)$ is independent of $s$ and equal to the cost function $C(\us)$ to be minimized, and $\beta(s)$ increases from 0 to $\infty$ as $s$ goes from 0 to 1. Note that $\beta$ and $E$ always appear through the combination $\beta E$; to simplify the discussion below we shall assume that $\beta$ is fixed to some arbitrary value and that all the time-dependence of the annealing schedule is included in $E(\us;s)$. We also assume that the initial condition at $s=0$ is at equilibrium.

We shall further assume that the dynamics flips one spin at a time, $W$ is hence non-zero only if $\us$ and $\us'$ are at Hamming distance at most 1. For definiteness we further specialize this choice to
\beq
W(\us^{(i)},\us;s) = \frac{1}{2}\left(1-\tanh\left(\beta \frac{E(\us^{(i)};s)-E(\us;s)}{2}\right)\right) \ .
\eeq
This form of the transition rate corresponds to the Glauber (or heat bath) rule, and can be expressed in terms of the effective magnetic field $h_i$ acting on spin $i$, defined as
\beq
h_i(\us;s) = \s_i \frac{E(\us^{(i)};s)-E(\us;s)}{2} \ .
\label{eq_def_heff}
\eeq
This expression is actually independent of $\s_i$.

\subsubsection{Quantum annealing}

This case corresponds to the evolution of an isolated system. Its wave function $|\psi(s)\ra$ thus follows Schr\"odinger's equation,
\beq
\frac{i}{\T} \frac{\dd  }{\dd s} |\psi(s)\ra = \hH(s) |\psi(s)\ra \ .
\eeq
We shall assume that the interpolation starts from a purely transverse field, $E(\us;s=0)=0$, with $|\psi(s=0)\ra$ the groundstate of the transverse field (i.e.\ the uniform superposition of the $2^N$ states of the computational basis), and that at the end of the annealing the transverse field is turned off and the classical energy corresponds to the cost function to be optimized: $\Gamma(s=1)=0$ and $E(\us;s=1)=C(\us)$.

\subsubsection{Simulated quantum annealing}

Quantum models can be represented as classical models with an additional imaginary time dimension (using for instance the Suzuki-Trotter identity~\cite{suzuki76}), of length corresponding to the inverse temperature. In the case of Ising spins in a transverse field one is thus led to consider a configuration space of trajectories, that we shall denote $\ubs=(\bs_1,\dots,\bs_N)$, with the convention that bold font symbols stand for imaginary-time dependent quantities, here $\bs_i =\{\s_i(t) : t \in[0,\beta] \}$ with $\s_i(t)=\pm 1$ is the imaginary time trajectory of an Ising spin. The quantum partition function and the thermodynamic average can then be computed by introducing the following equilibrium measure on the space of trajectory configurations (for a classical energy $E(\us)$, a transverse field $\Gamma$ and an inverse temperature $\beta$):
\beq
P_{\rm eq}(\ubs) = \frac{1}{Z} \prod_{i=1}^N \Gamma^{|\bs_i|} e^{-\int_0^\beta E(\us(t)) \dd t} \ ,
\label{eq_Peq_quantum}
\eeq
where $|\bs_i|$ denotes the number of spin flips in the trajectory of the $i$-th spin, and we assume periodic boundary conditions in the imaginary time direction, $\us(t=0)=\us(t=\beta)$. A Quantum Monte Carlo simulation proceeds by a stochastic exploration of the space of configurations $\ubs$. One can use for instance a quantum version of the heat-bath algorithm, where elementary moves correspond to the update of one spin trajectory $\bs_i$ (for the whole imaginary time axis), redrawn from the equilibrium measure conditioned on the current state of all trajectories on sites distinct from $i$. What we call simulated quantum annealing is precisely this type of stochastic process for $\ubs$, described by the master equation,
\beq
\frac{1}{\T} \frac{\dd  }{\dd s} P(\ubs;s) = \sum_{\ubs'} W(\ubs,\ubs';s) P(\ubs';s) \ ,
\eeq
where the transition rates are those of heat-bath updates, respecting detailed balance with respect to the instantaneous equilibrium distribution (\ref{eq_Peq_quantum}) with the current values of $E(\cdot;s)$ and $\Gamma(s)$ (for simplicity of notation we assume the inverse temperature $\beta(s)$ to be held constant). The initial condition will be considered to be equilibrated at $s=0$, i.e.\ in a purely transverse field.

\subsection{Mean-field fully-connected models}

The models we shall consider are generalizations of the (quantum) Curie-Weiss model (also related to the Lipkin-Meshkov-Glick model~\cite{LMG}), with their energy depending on the configurations only through the total magnetization:
\beq
E(\us;s)=N g\left(\frac{1}{N}\sum_{i=1}^N \s_i;s\right) \ ,
\eeq
$g(m;s)$ being the energy density for a configuration of magnetization density $m$ at reduced time $s$. The effective field (see the definition in Eq.~(\ref{eq_def_heff})) created by such a configuration then reads
\beq
h(m;s) = - \frac{\partial}{\partial m} g(m;s) \ ,
\eeq
at the leading order in the large $N$ limit.

As the transverse field $\Gamma(s)$ is also conjugated to the total (transverse) magnetization, such quantum models are invariant under any permutation of the spins (or in more physical terms their Hamiltonians commute with the total spin), which greatly simplifies their analysis.

\section{Evolution equations}
\label{sec_evolution_eq}
\subsection{Differential equations}

As we shall see now the high level of symmetry in the definition of these models allows one to reduce the study of their dynamics, in the thermodynamic limit $N\to\infty$ with the annealing time $\T$ kept finite, to simple differential equations, for the three type of dynamics introduced above and arbitrary time dependencies of their parameters.

\subsubsection{Thermal annealing}
If the initial condition $P(\us;s=0)$ of the master equation (\ref{eq_master_thermal}) is invariant under the permutation of the sites (i.e.\ only depends on the magnetization of the initial condition), then this property is conserved along the dynamical evolution (even for finite $N$), allowing one to deduce a master equation for $P(m;s)$. In addition in the thermodynamic limit this probability distribution concentrates around its most likely value $m(s)$ (if the initial condition is not too wide), the latter being solution of a simple, deterministic, differential equation:
\beq
\frac{1}{\T} \frac{\dd m}{\dd s} = \tanh(\beta h(m(s);s)) - m(s) \ .
\label{eq_m_thermal}
\eeq
This expression, that can be easily obtained after a short computation, has a natural interpretation: when a spin is updated its average magnetization corresponds to that of a single spin at equilibrium at inverse temperature $\beta$, in an effective magnetic field created by the rest of the configuration and thus equal to $h(m(s);s)$.

\subsubsection{Quantum annealing}

The Schr\"odinger evolution of such mean-field spin models was studied in details in~\cite{BaSe12} (see also~\cite{sciolla11,ScFa10} for similar treatments of particle models), where it was shown that the thermodynamic limit leads to semi-classical Hamiltonian equation of motions for the typical magnetization $m(s)$ and its conjugated momentum $\tm(s)$:
\beq
\begin{cases} 
\frac{1}{\T} \frac{\dd}{\dd s}  m(s)  =
\frac{\partial}{\partial \tm} \mathcal{H}(m(s),\tm(s);s)  & \\
\frac{1}{\T} \frac{\dd}{\dd s} \tm(s)  = -  
\frac{\partial}{\partial m} \mathcal{H}(m(s),\tm(s);s) &
\end{cases}
\ ,
\label{eq_hamilton}
\eeq
where the effective Hamiltonian $\mathcal{H}$ is given in terms of the parameters of the model as
\beq 
\mathcal{H}(m,\tm;s) = g(m;s)-\Gamma(s)\sqrt{1-m^2} \cos(2 \tm) \ .
\eeq
The initial condition is $(m(s=0),\tm(s=0))=(0,0)$.

\subsubsection{Simulated quantum annealing}

Let us finally consider the case of the annealing of a Quantum Monte Carlo simulation. Its analysis is at first sight much more complicated than the thermal annealing one, as the configurations have an additional imaginary time dependence. It turns out however that in the large $N$ limit the magnetization density concentrates around a typical value $m(s)$, independently of the imaginary time (i.e.\ the static approximation of~\cite{Inoue} becomes exact), and that the correlations along this axis are irrelevant (see~\cite{BaSe13long} for a detailed derivation of this statement). This magnetization density is found to obey a simple differential equation,
\beq
\frac{1}{\T} \frac{\dd m}{\dd s} = \frac{h(m(s);s)}{\sqrt{h(m(s);s)^2 + \Gamma(s)^2}} \tanh(\beta \sqrt{h(m(s);s)^2 + \Gamma(s)^2} ) - m(s) \ ,
\label{eq_m_squantum}
\eeq
as was already obtained in~\cite{Inoue} for a particular choice of $h(m,s)$. As it should this equation reduces to (\ref{eq_m_thermal}) when the transverse field $\Gamma(s)$ vanishes at all times. In fact the intuitive explanation given after (\ref{eq_m_thermal}) remains valid here, as the first term in (\ref{eq_m_squantum}) is precisely the average value of the $z$-magnetization of a quantum spin $1/2$ in a longitudinal field $h$ and transverse field $\Gamma$.

\subsection{Analysis in the large annealing time limit}

For any finite value of $\T$ the evolution equations (\ref{eq_m_thermal}), (\ref{eq_hamilton}) and (\ref{eq_m_squantum}) can be easily integrated numerically, for any annealing schedule. One can however make further analytical progresses by considering the limit $\T\to\infty$ of a very slow annealing (we recall that we have already taken the thermodynamic limit $N\to\infty$). As the thermal annealing is a particular case of the simulated quantum annealing we shall not discuss explicitly the thermal case, and contrast the behavior of the quantum and simulated quantum annealing.

Consider the evolution equation (\ref{eq_m_squantum}) of the latter. In the large $\T$ limit one can check that $m(s)$ will be close to $m_*(s)$, where $m_*(s)$ is such that the right-hand side of (\ref{eq_m_squantum}) vanishes. This condition is easily seen to be equivalent to $m_*(s)$ being an extremum of the free energy $f(m;s)$, a standard computation (see for example~\cite{BaSe12}) leading indeed to
\beq
f(m;s) =\underset{\lambda}{\text{ext}}  \left[ 
g(m;s) + m \lambda - \frac{1}{\beta} \ln 2 \cosh( \beta \sqrt{(1-s)^2 +\lambda^2}) \right] \ . 
\eeq
Of course the free energy can have several extrema, in that case one has to be more precise to define $m_*(s)$. The rules to remove this ambiguity are as follows. $m_*(0)$ has to be chosen as the local minimum reached by a steepest-descent from the initial condition. Then $m_*(s)$ remains stuck in the same local minimum as long as the latter exists. At all bifurcations where $m_*(s)$ loses its stability, it flows to a new local minimum following a steepest-descent search of the free energy. For future use let us write down a more explicit form of the above free energy in the zero-temperature limit:
\beq
f(m;s) = g(m;s) - (1-s) \sqrt{1-m^2} \ .
\label{eq_f_zero_T} 
\eeq

\begin{figure}
\centerline{\includegraphics[width=10cm]{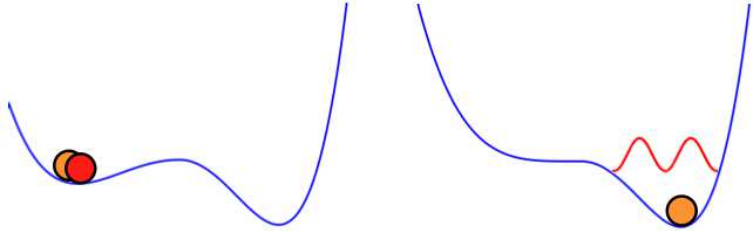}}
\caption{Sketch of the difference of behavior between the zero temperature simulated quantum annealing (orange circle) and the quantum annealing (red circle and wavelet) when crossing a spinodal point along their annealing.}
\label{fig_sketch_spinodal}
\end{figure}

Let us now turn to the analysis of the slow annealing limit of the quantum annealing equations~(\ref{eq_hamilton}). These have the form of Hamiltonian equations of motion, and their large $\T$ limit can thus be analyzed using the theory of adiabatic invariants of classical mechanics. The initial condition $(m(s=0),\tm(s=0))=(0,0)$ is a local minimum of the Hamiltonian $\mathcal{H}(m,\tm;s=0)$, hence for large annealing times $\T$ the solution of (\ref{eq_hamilton}) remains close to $(m(s),\tm(s))=(m_*(s),0)$, as long as this point corresponds to a local minimum of $\mathcal{H}$. This is actually also the minimum of the zero-temperature free energy (\ref{eq_f_zero_T}), hence the quantum annealing and the zero-temperature simulated quantum annealing coincide in this limit as long as no bifurcation is encountered. However a drastic difference between the two dynamics appears as soon as the local minimum loses its stability: the semi-classical dynamics describing the quantum annealing is indeed conservative, hence a non-zero value of the momentum $\tm$ is induced at the bifurcation, and an oscillatory behavior for $m(s)$ in its potential energy well takes place. On the contrary the zero-temperature simulated quantum annealing flows to the newly accessible local minimum. This difference of behavior is illustrated on figure~\ref{fig_sketch_spinodal}.

\section{Conclusions}
\label{sec_conclusions}

In this paper we have studied various types of annealing dynamics of quantum spin models, simple enough to be reducible to unidimensional problems, their magnetization playing the role of a particle coordinate. The main message to be emphasized is the difference between a quantum annealing (an evolution with Schr\"odinger equation) and a low temperature simulated quantum annealing (a time-varying Quantum Monte Carlo simulation) that appears when a bifurcation occurs in the structure of the effective energy potential along the annealing (see also~\cite{hastings_obstructions} for a discussion of this difference from another point of view). This difference persists even in the limit of zero-temperature, which can be understood physically as a system coupled to a ``zero-temperature heat bath'' can release part of its energy, at variance with an isolated system. Microscopic descriptions of the coupling to the environment might thus be necessary to interpret the results of experiments on quantum annealers~\cite{dwave_troyer}.

When the groundstate of such a model encounters a first-order quantum phase transition both quantum and simulated quantum annealing remains stuck in the metastable continuation of the groundstate between the transition and the spinodal point. Right after the spinodal the energy is strictly smaller in the simulated quantum annealing case, as the coupling to the heat bath provides a mechanism for energy relaxation. This could seem to suggest that including both thermal and quantum fluctuations is the best strategy to attain the lowest possible residual energy. This conclusion is however wrong in general, as the subsequent relative evolution of the depth and width of the multiple energy wells between the spinodal and the end of the annealing can in some cases reverse this ordering~\cite{BaSe13long}.

The long-term goal sketched in the introduction would be to understand these annealing dynamics in disordered mean-field systems that correspond to random optimization problems. The discussion presented here in terms of effective free energy is reminiscent of the quantum and dynamical TAP equations approach of~\cite{Bi99,BiCu01} for disordered fully-connected models. Such models, even if obviously much more intricate than the simple ones discussed here, share with the latter the possibility of first-order phase transitions accompanied by metastable states and spinodals (thanks to their mean-field character), one can thus hope for some level of universality for the mechanism encountered at spinodals. As an intermediate step in this program we shall present in~\cite{BaSe13long} further results on the quantum random subcubes model~\cite{rcm,foini10}; in the regime where the number of cubes is finite one can exploit the permutation symmetry between sites inside a few blocks of spins, which leads to a similar analysis of the annealing dynamics with the magnetization replaced by a finite dimensional vector of block magnetizations.

Let us finally emphasize that we concentrated here on a particular time regime, namely with the limit $\T \to \infty$ of long annealing times taken after the thermodynamic limit $N\to\infty$. In presence of metastability this time scale is however too short to reach the groundstate energy, the adiabatic time (for quantum or simulated annealing) growing exponentially fast with the system size. It is thus also important to consider the regime where $\T$ and $N$ are taken to infinity simultaneously, with $\T=\exp[N \tau]$. We have shown in~\cite{BaSe12} how to compute the residual energy of the quantum annealing as a function of $\tau$ via the determination of the exponentially small avoided level crossings along the metastable branches of the eigenvalues of the Hamiltonian $\hH(s)$. In the case of simulated quantum annealing one can also investigate this exponential time regime by computing the height of the free energy barriers, the adiabatic time being controlled by these barriers through an Arrhenius law~\cite{BaSe13long}.

\end{document}